# X-ray parametric down-conversion probes radiation-induced dynamics in hybrid perovskite $MAPbBr_3$

S. Sofer[1,2], O. Sefi[1,2], G. J. Man[1,2*], H.N. Barad[1,3], N.P. Jasti[3,4], D. Cahen[4], and S. Shwartz[1]

[1] *Inst. of Nanotechnology & Adv. Mater. (BINA), Bar-Ilan Univ., Ramat-Gan 52900 Israel*

[2] *Dept. of Physics, Bar-Ilan University, Ramat Gan, 52900 Israel*

[3] *Dept. of Chemistry, Bar-Ilan University, Ramat Gan, 52900 Israel*

[4]*Dept. of Molec. Chem. & Mater. Sci., Weizmann Inst. of Science, Rehovoth 76100, Israel.*

* *Currently at Gabriel Man och Partners AB, Uppsala 752 57, Sweden and Man Labs Technologies LLC, Dallas, Texas 75219, USA*

We report minute-timescale time-resolved measurements of X-ray parametric down-conversion (PDC) of X-rays into ultraviolet photons in a $MAPbBr_3$ single crystal under continuous irradiation. Despite the known radiation sensitivity of hybrid perovskites and the high photon flux required to detect the intrinsically weak PDC signal, the nonlinear response remains measurable throughout the irradiation sequence. Both PDC and Bragg diffraction exhibit a slow decay consistent with diffusion-driven damage dynamics on timescales reported for bromide-vacancy and proton migration. However, despite their similar decay rates, the PDC and Bragg signals evolve toward markedly different steady states, while an additional fast decay component is observed in the Bragg response but absent in the PDC signal. Density functional theory calculations identify N 2s states of the methylammonium cation as the most likely contributors to the measured PDC response and show that their Fourier component is only weakly affected by bromide-vacancy formation. These results establish nonlinear X-ray diffraction as a channel-selective probe of radiation-induced dynamics and demonstrate that radiation damage can couple differently to nonlinear and linear X-ray responses.

X-ray/optical wave-mixing processes provide direct access to reciprocal-space Fourier components of the optical response at the generated frequency via phase-matching with crystal reciprocal lattice vectors [1-16]. Whereas X-ray Bragg diffraction measures structural Fourier components of the crystal through the electron density, x-ray/optical wave mixing measures Fourier components of its optical response. Consequently, it can access phenomena that are inaccessible to diffraction alone, including microscopic valence electronic structure [7,12], local symmetry breaking [13], and the reorganization of unoccupied electronic states across quantum phase transitions [14].

Among these techniques, X-ray parametric down-conversion (PDC) provides a particularly direct implementation in which an incident X-ray converts into a correlated X-ray/UV photon pair through interaction with vacuum fluctuations. The generated UV photon interacts with the material, while the correlated x-ray photon emerges from the crystal carrying information about the optical response at the UV frequency. Unlike other X-ray/optical wave-mixing experiments [4,15,16], no external laser is required, enabling access to Fourier components of the UV optical response without perturbing the system through external excitation.

However, PDC is intrinsically weak and typically requires long acquisition times and high X-ray fluxes. Consequently, previous studies have focused on systems that remain effectively static throughout the measurement [5-11,14]. Whether PDC can provide meaningful time-resolved information from radiation-sensitive materials, where the probing beam itself drives structural and electronic evolution, remains an open question. More importantly, it is unknown whether the optical-response Fourier components accessed by PDC provide information beyond that available from conventional X-ray diffraction under such conditions. Addressing these questions is essential for establishing PDC as a probe of time-dependent phenomena in radiation-sensitive materials.

Hybrid organic–inorganic perovskites are a prominent class of semiconductors whose lattice is chemically susceptible to ionizing radiation [17,18]. Owing to their high absorption coefficients, long carrier diffusion lengths, and low-cost synthesis, they have attracted considerable interest in optoelectronic applications such as photovoltaics, light emission, and radiation detection [19-21]. Among them, methylammonium lead bromide ($MAPbBr_3$) single crystals have emerged as promising X-ray detector materials due to their high sensitivity and efficient charge collection [22,23]. Perovskite photovoltaics, X-ray detectors, and light-emitting diodes are progressing toward deployment. However, the mechanisms of photo- and radiation-induced degradation remain incompletely understood [17,18,24,25], which may limit device lifetime in the field.

X-ray irradiation of halide perovskites is known to induce a complex interplay of ionic migration, radiolytic decomposition, and structural rearrangement [17,18,26]. Results from previous studies have been interpreted as formation and migration of halide vacancies [27,28], while decomposition of the organic cation [24,25], and the release of volatile degradation products, including HBr and $CH_3NH_2$ [25,29], were directly confirmed experimentally. The coexistence of these processes makes radiation damage in halide perovskites inherently multifaceted and motivates the development of probes sensitive to complementary aspects of the evolving material.

Here, we present time-resolved measurements of both PDC and Bragg intensities in a $MAPbBr_3$ single crystal under continuous X-ray irradiation. We show that non-linear X-ray diffraction can be measured in a radiation-sensitive material as it evolves under irradiation, providing complementary information to that from conventional diffraction. We use density functional theory calculations show that the PDC decay cannot be explained by bromide migration and its effect on the induced charge density. Together, these results establish nonlinear X-ray diffraction as a real-time probe of radiation-induced dynamics and show that the same radiation-induced processes can couple differently to microscopic optical and structural observables.

The experiment was performed at beamline I16 of Diamond Light Source [30] using a monochromatic 10 keV X-ray beam selected by a Si(111) monochromator. The experimental arrangement is illustrated in Fig. 1. A $MAPbBr_3$ single crystal, grown by inverse temperature crystallization from a stoichiometric (1:1) $MABr:PbBr_2$ solution in DMF [31], was mounted on the six-circle diffractometer and oriented to access the (3,0,0) reciprocal lattice vector, which provides a convenient phase-matching geometry for the PDC measurements. The measurements were carried out under vacuum at ambient room temperature. Switching between the Bragg and PDC conditions required only a slight crystal rotation, while the detector position remained fixed because its angular acceptance captured both signals. The detector was a Quad-Merlin area detector with dimensions of about 28 × 28 mm and pixel size of 55 μm × 55 μm (512 × 512 pixels). The scattering angle for each pixel was calculated from its detector position and the known sample-to-detector distance.

The primary experimental challenge in measuring X-ray PDC is separating the weak nonlinear signal from elastic scattering that is several orders of magnitude more intense. This is achieved by using a three-bounce Si(111) crystal analyzer positioned between the sample and the QuadMerlin detector, providing an energy resolution of approximately 1 eV at 10 keV. The analyzer functions as a narrow-band energy filter, rejecting photons at the pump energy while transmitting the energy-shifted PDC signal. Throughout this work, the idler energy was fixed at 40 eV by tuning the crystal analyzer to the corresponding signal-photon energy. This energy was selected because it probes shallow core and valence electronic states while providing sufficient nonlinear signal for the experiment. Here, the idler denotes the lower-energy UV photon, while the signal denotes the higher-energy X-ray photon. The incident beam was defined by a 50 × 50 μm slit aperture clipping a naturally focused beam of ~ 200 μm (horizontal) × 30 μm (vertical), reducing the incoming flux (~$10^{13}$ photons/s) roughly 4x. The sample was mounted at 1310 mm from the detector and 1250 mm from the analyzer. Due to the known sensitivity of halide perovskites to radiation, each time-resolved measurement sequence was initiated at a previously unexposed location on the crystal. At the start of each measurement, we probed a spot on the crystal that was not exposed to radiation (verified by the initial intensity of the Bragg peak).

To verify that the detected signal arises from PDC, we measured rocking curves for several detector angles and compared the results with the phase-matching condition $\overrightarrow{k_s} + \overrightarrow{k_\iota} = \overrightarrow{k_p} + \vec{G}$ where $\overrightarrow{k_p}, \overrightarrow{k_s}, \overrightarrow{k_\iota}$ are the wave vectors for the pump, signal and idler photons, respectively, and $\vec{G}$ is the reciprocal lattice vector. For each rocking curve, the background was determined as the mean count rate measured at an off-peak position over the region of interest on the detector, far from the PDC peak, and subtracted from the raw intensity. The resulting rocking curves in Fig. 2a show a clear peak at each detector angle, with the peak position shifting systematically as expected from momentum conservation. Figure 2(b) compares the measured peak positions with the phase-matching angles calculated for G = (3,0,0). Slight deviations from phase matching could arise from uncertainties in the refractive indices of the three photons. The overall agreement confirms that the observed signal arises from PDC.

Having established the parametric origin of the detected signal, we next investigated its evolution under continuous X-ray irradiation. Bragg and PDC rocking curves were acquired sequentially throughout the exposure, allowing the temporal evolution of both the structural and optical responses associated with the (3,0,0) reciprocal lattice vector to be tracked under identical irradiation conditions. Each point was recorded at intervals of ~45-minutes. Each point in Fig. 3 corresponds to the peak of a rocking curve, measured with a counting time of 1 s for the Bragg signal and 10 s for the PDC signal.

Figure 3 presents the measured Bragg (3,0,0) and PDC integrated intensities as a function of exposure time. Successive Bragg and PDC pairs were recorded at intervals of ~46 min (2780 s), for a total exposure of ~10.8 h. Within each pair, the PDC rocking curve was acquired about 3.5 minutes after the Bragg rocking curve. The Bragg intensity is plotted on a logarithmic scale to capture its large dynamic range, whereas the PDC intensity is shown on a linear scale. Both signals decay over the course of the measurement, but with visibly different behavior. The Bragg intensity decreases rapidly within the first few thousand seconds, losing roughly an order of magnitude before transitioning to a slower decay. In contrast, the PDC signal decays more gradually and uniformly throughout the exposure, without an obvious fast initial component. The distinct temporal evolution of the two signals already indicates that the irradiation-induced structural changes probed directly by Bragg diffraction are only weakly coupled to the evolution of the electronic structure probed indirectly through the PDC intensity, which depends on the occupation of the initial and final electronic states as well as the corresponding transition dipole moment. The rapid initial decrease of the Bragg intensity therefore points to a damage process that affects the crystal lattice while having only a limited influence on the electronic transitions governing the PDC signal.

To quantify the differences between the Bragg and PDC decay dynamics, the measured intensity traces were fitted using simple models for exponential decays. The Bragg intensity is modeled as a bi-exponential decay of the form

$$I_{Bragg}(t) = A_1 \, exp(-t/\tau_1) + A_2 \, exp(-t/\tau_2) + C \tag{1}$$

where $A_1$ and $A_2$ are the amplitudes of the fast and slow components, $\tau_1$ and $\tau_2$ are the corresponding time constants, and C is a constant offset accounting for residual scattering at long times. All values are fitted to the background removed measurements. The fit yields $\tau_1 = 832 \pm 93$ s and $\tau_2 = 9656 \pm 2140$ s, with the fast component dominating the initial decay and the slow component governing the long-time behavior. A single exponential model fails to reproduce the steep initial decline, indicating that the structural dynamics measured by Bragg diffraction are governed by two distinct processes.

The PDC intensity is adequately described by a single-exponential decay,

$$I_{PDC}(t) = A\,exp(-t/\tau_{PDC}) + C \tag{2}$$

yielding $\tau_{PDC}$ = 8827±1070 s. This value is in close quantitative agreement with the slow Bragg component $\tau_2$. The absence of a fast component in the PDC decay indicates that the process responsible for the rapid early loss of Bragg intensity couples only weakly to the optical-response Fourier component probed at an idler energy of 40 eV. The close correspondence between the PDC and slow Bragg time constants, nevertheless, suggests a common mechanism underlying the long-time evolution of both signals.

To connect the measured time constants to microscopic transport, we obtain order-of-magnitude estimates of the effective diffusion coefficients using the relation [32]:

$$D = L^2/\tau, \tag{3}$$

where L = 32.3 μm is the calculated X-ray absorption length of $MAPbBr_3$ at 10 keV [33]. We use the absorption length as the characteristic transport length because X-ray attenuation produces a depth-dependent rate of defect generation over this spatial scale. The extracted values are summarized in Table 1. The diffusion coefficients obtained from the slow Bragg and PDC, $D_2$ = (1.08± 0.24) × $10^{-9}$ cm²/s and $D_{PDC}$ = (1.18± 0.14) × $10^{-9}$ cm²/s, respectively, are nearly identical. Both values fall within the range of $10^{-9}$ to $10^{-10}$ cm²/s assigned to proton diffusion through the halide sublattice in hybrid perovskite single crystals [34]. The close correspondence between the PDC and slow Bragg diffusion coefficients therefore supports the interpretation that the long-time evolution of both signals is governed by a common transport process. We note that diffusion coefficients of similar magnitude have also been assigned to bromide-vacancy migration [27,28] although these assignments rest mainly on the magnitude of the diffusion coefficients rather than on direct identification of the mobile species. The DFT calculations presented below provide further insight into the possible microscopic origin of the PDC decay. They identify the N 2s states of the methylammonium cation as the most likely contributors to the 40 eV PDC channel and show that their Fourier component is only weakly affected by bromide-vacancy formation. In contrast, deprotonation of the cation directly modifies the nitrogen bonding environment and could therefore perturb these states. This makes proton-related defects a plausible contributor to the observed PDC decay.

Of importance**,** the Bragg reflection retains only 1.5% of its initial intensity at long times, while the PDC signal saturates when it reaches 8.6% of the initial intensity. This disparity is consistent with Bragg probing the extinction length [35], compared to the absorption length sampled by PDC, where deeper regions receive a lower cumulative dose due to beam attenuation.

The fast Bragg component yields $D_1 \approx 1.25 \times 10^{-8}$ cm²/s, more than an order of magnitude above the diffusion coefficients associated with either $Br^-$ vacancy or proton migration in $MAPbBr_3$. This value is therefore inconsistent with the transport process responsible for the slow Bragg and PDC decays and suggests a distinct damage mechanism.

A natural candidate is radiolytic degradation. Under intense X-ray irradiation, both the organic cation and the lead-halide framework can undergo bond scission, leading to the formation of metallic lead, halide desorption, and loss of volatile organic species [17,18]. The fluence scale of the fast Bragg decay is quantitatively consistent with this interpretation: a 20% loss of Bragg intensity occurs at $t \approx 209$ s, corresponding to a cumulative fluence of $(3.8–7.7) \times 10^{17}$ ph/cm², within a factor of two of the fluence at which comparable degradation was observed in photoelectron spectroscopy on $MAPbBr_3$ single crystals [17].

The absence of a corresponding fast component in the PDC decay suggests that the radiolytic process responsible for the rapid initial loss of Bragg intensity contributes only weakly to the measured PDC signal. One possible explanation lies in the different sampling depths of the two measurements. Bragg diffraction probes a shallow near-surface region characterized by the extinction length, the characteristic distance over which the incident x-ray beam transfers a substantial fraction of its energy to the diffracted wave through dynamical diffraction. Under the present experimental conditions, this length is estimated to be only a few micrometers. In contrast, because PDC is an intrinsically weak process, the detected nonlinear signal is generated over a substantially larger crystal volume, with an effective sampling depth comparable to the 10 keV x-ray absorption length of approximately 30 μm. Bragg diffraction is therefore more sensitive to damage near the crystal surface, where the absorbed dose is highest, whereas a thin damaged surface layer contributes only a small fraction of the integrated PDC signal. In this picture, the slower diffusion-mediated process affecting a larger crystal volume dominates the observed PDC dynamics.

To gain further insight into the origin of the observed decay dynamics, we examine the evolution of the rocking-curve widths and peak positions. Figure 4 shows how the measured full width at half maximum (FWHM) and peak position of the Bragg (3,0,0) and PDC signals evolve with exposure time.

The FWHM evolution (Fig. 4a) reveals strikingly different behavior for the two signals. The Bragg FWHM increases monotonically from approximately 0.012° at the start of the exposure, rising steeply over the first 5000 s before gradually saturating near 0.035° for the remainder of the measurement. In contrast, the PDC FWHM follows a non-monotonic trajectory. Starting from approximately 0.049° it increases to a maximum of approximately 0.073° around 7000–8000 s before reversing and decreasing steadily to approximately 0.039° by the end of the exposure. This contrasting behavior reflects the different dependence of the Bragg and PDC rocking curves on the evolving crystal properties and phase-matching conditions.

The peak-position evolution (Fig. 4b) provides a further distinction between the two signals. The Bragg peak drifts continuously to lower angles over the full measurement, shifting from 17.985° to approximately 17.953 ± 0.002°, corresponding to a total displacement of about 0.032° which translates to an interplanar spacing modulation of ~ 0.3 pm. In contrast, the PDC peak position remains essentially constant near 18.00° throughout the exposure, with fluctuations remaining within the confidence intervals of the Gaussian fits. The combination of a shared slow decay timescale with markedly different peak-position evolution is consistent with that PDC and Bragg diffraction providing complementary and non-redundant views of the radiation damage dynamics.

To understand the microscopic origin of this complementary sensitivity, we next examine the electronic states contributing to the measured PDC channel. The two transport mechanisms consistent with the observed diffusion coefficient couple to the PDC signal in different ways. Bromide-vacancy formation removes an ion together with its electrons and can therefore directly modify the induced charge density at the measured idler energy. A migrating proton, in contrast, carries no electronic states and has no resonance at 40 eV, so it can affect the PDC signal only indirectly, through changes in the electronic structure of the deprotonated methylammonium cation. We therefore first test the direct mechanism and ask whether the 40 eV idler signal is sensitive to bromide-vacancy formation. The measured PDC intensity is governed by the nonlinear susceptibility $\chi_{\vec{G}}^{(2)}(\omega_s)$. In a crystal, the nonlinear susceptibility can be related to the Fourier component of the linear susceptibility at the idler frequency, $\chi_{\vec{G}}^{(1)}(\omega_i)$, through [7]:

$$\chi_{\vec{G}}^{(1)}(\omega_i) = \frac{2mc\omega_s}{e\theta_{psi}} \chi_{\vec{G}}^{(2)}(\omega_s), \tag{4}$$

where $\omega_s$ is the frequency of the signal photon, $\theta_{psi}$ is the polarization factor, $m$ is the electron mass, $c$ is the speed of light and $e$ is the electron charge. This relation shows that the nonlinear PDC signal, $\chi_{\vec{G}}^{(2)}$, provides direct access to the linear electronic response of the crystal at the idler energy, $\chi_{\vec{G}}^{(1)}$. Consequently, the measured PDC efficiency is directly sensitive to the Fourier component of the induced charge density at 40 eV, providing the basis for the electronic-structure analysis presented below. To determine whether the PDC signal is directly linked to bromide-vacancy formation, we performed density functional theory calculations using Quantum ESPRESSO v7.1 [36,37] with projector-augmented-wave (PAW) pseudopotentials, the Perdew–Burke–Ernzerhof (PBE) exchange-correlation functional [38], and Grimme-D3 dispersion corrections [39]. A 2×2×2 supercell of cubic $MAPbBr_3$ (96 atoms, a = 11.839 Å) was constructed and sampled with a 2×2×2 k-point mesh. P1 symmetry was adopted because the $MA^+$ cation's fixed orientation already lowers the nominal cubic symmetry, and the Br vacancy introduces further local distortion upon relaxation. Using P1 for both pristine and defective cells avoids imposing artificial constraints on either. The bromide-vacancy configuration (95 atoms) was obtained by removing one Br ion and relaxing all internal coordinates at fixed cell parameters. Within the Kohn-Sham eigenvalue spectrum, the states nearest to 40 eV binding energy are the N 2s levels of the methylammonium cation, appearing at approximately −17.3 eV absolute (the systematic underestimate relative to the experimental value is a well-known limitation of semi-local DFT [38] and does not affect the orbital identification). The Br 4s states, localized on the inorganic framework, lie in a separate energy window around −11 to −10 eV absolute. This energetic separation suggests that the electronic transitions involving the N 2s states of the methylammonium cation are the most likely contributors to the response probed by the 40 eV PDC channel.

This observation, however, does not rule out the possibility that the 40 eV PDC channel is indirectly sensitive to bromide-vacancy formation. To assess how bromide-vacancy formation affects the electronic states most likely contributing to the 40 eV PDC response, we compare their orbital-resolved charge densities and Fourier components at the measured reciprocal lattice vector.

Figures 5(a) and 5(b) show that the charge density remains tightly localized around the nitrogen atoms of the $MA^+$ cations and is essentially unaffected by the removal of a bromide ion from the inorganic sublattice.

This observation is supported quantitatively by the results in Table 2**,** which presents the Fourier amplitude of the electron density $|\rho_{3,0,0}|$ for pristine and bromide vacancy supercells. For the N 2s states, the Fourier amplitude decreases by only 1.8% upon vacancy formation, from $2.82\times10^{-3}$ to $2.77\times10^{-3}$ $e\cdot Å^{-3}$. This weak response is expected because the N 2s orbital is a deep, atomic-like state confined within the organic cation, with negligible spatial overlap with the Br sites.

By contrast, Figs. 5c,d show that the Br 4s states respond directly to the vacancy through the disappearance of charge density at the missing Br site. The corresponding Fourier amplitude changes by 52.2%, from $4.52\times10^{-4}$ to $2.16\times10^{-4}$ $e\cdot Å^{-3}$. The large contrast between the N 2s and Br 4s responses demonstrates that the methodology is sensitive to bromide-vacancy formation when the relevant electronic states reside on the affected sublattice. The weak response of the N 2s Fourier component indicates that bromide-vacancy formation only weakly perturbs the electronic states most likely contributing to the 40 eV PDC response.

The DFT results therefore strongly disfavor a simple interpretation in which the measured PDC decay directly tracks the bromide-vacancy concentration. Yet the PDC and slow Bragg signals exhibit nearly identical characteristic timescales. This combination of observations suggests that the measured PDC response is coupled only indirectly to the diffusion-mediated defect dynamics. One possible explanation is that the slow defect dynamics are accompanied by broader chemical or structural modifications that alter the local environment of the methylammonium cations and, consequently, the N 2s-derived electronic states contributing to the 40 eV channel. A specific candidate for such a modification is proton migration, which is consistent with the same diffusion coefficient.

Unlike halide vacancies, proton migration cannot be assessed directly from the induced charge density at 40 eV. The migrating proton carries no electronic states and therefore has no resonance at this idler energy. However, proton migration originates from deprotonation of the methylammonium cation, which modifies the bonding environment of the nitrogen atom and is therefore expected to perturb the N 2s-derived states that dominate the 40 eV channel. Any coupling of proton migration to the PDC signal is thus indirect, mediated by changes in the electronic structure of the remaining $CH_3NH_2$ species rather than by the proton distribution itself. Our DFT calculations consider only bromide-vacancy formation, so they do not allow a quantitative conclusion on the proton-related contribution. The agreement between the PDC decay constant and the diffusion coefficient expected for proton migration, together with the weak sensitivity of the 40 eV channel to bromide vacancies, makes deprotonation-related defects a plausible origin of the observed PDC decay. A definitive assignment would require explicit calculations of the deprotonated configuration.

In conclusion, we have measured the time-resolved decay of X-ray PDC in $MAPbBr_3$ under continuous irradiation and compared it with the corresponding Bragg diffraction response. Despite the high radiation dose required to detect the weak nonlinear signal, the PDC response remains observable throughout the experiment and exhibits dynamics distinct from those measured by conventional diffraction. The PDC signal at 40 eV idler energy decays as a single exponential whose time constant matches the slow component of the bi-exponential Bragg decay, yielding a diffusion coefficient of $1.18 \times 10^{-9}$ $cm^2/s$, consistent with both bromide vacancy migration and proton migration through the halide sublattice. The fast Bragg component, with a diffusion coefficient an order of magnitude larger, is attributed to radiolysis of the organic cation in the near-surface layer and has no counterpart in the PDC response. Despite sharing a slow decay timescale, the PDC and Bragg signals diverge in their rocking-curve evolution. DFT calculations identify N 2s states of the methylammonium cation as the most likely contributors to the 40 eV PDC response and show that their charge-density Fourier component is only weakly affected by bromide-vacancy formation. The observed PDC decay therefore does not directly track the halide-vacancy concentration, indicating that the nonlinear response is coupled to the damage dynamics through a different electronic pathway. These results establish X-ray PDC as a real-time probe of radiation induced dynamics and demonstrate that optical-response and structural Fourier components can evolve differently under the same irradiation conditions. More broadly, they highlight the potential of nonlinear X-ray diffraction to provide information inaccessible to conventional diffraction through the selective probing of electronic-state Fourier components. Since the electronic states contributing to the nonlinear response are determined by the idler energy, different PDC channels may provide sensitivity to distinct microscopic radiation-induced processes.

This work was supported by the Israel Science Foundation (ISF) (IL), Grant No. 2208/24. This work was supported in part by the Zuckerman STEM Leadership Program (GJM). We acknowledge Diamond Light Source for the provision of time on Beamline I16 under Proposal 34685.

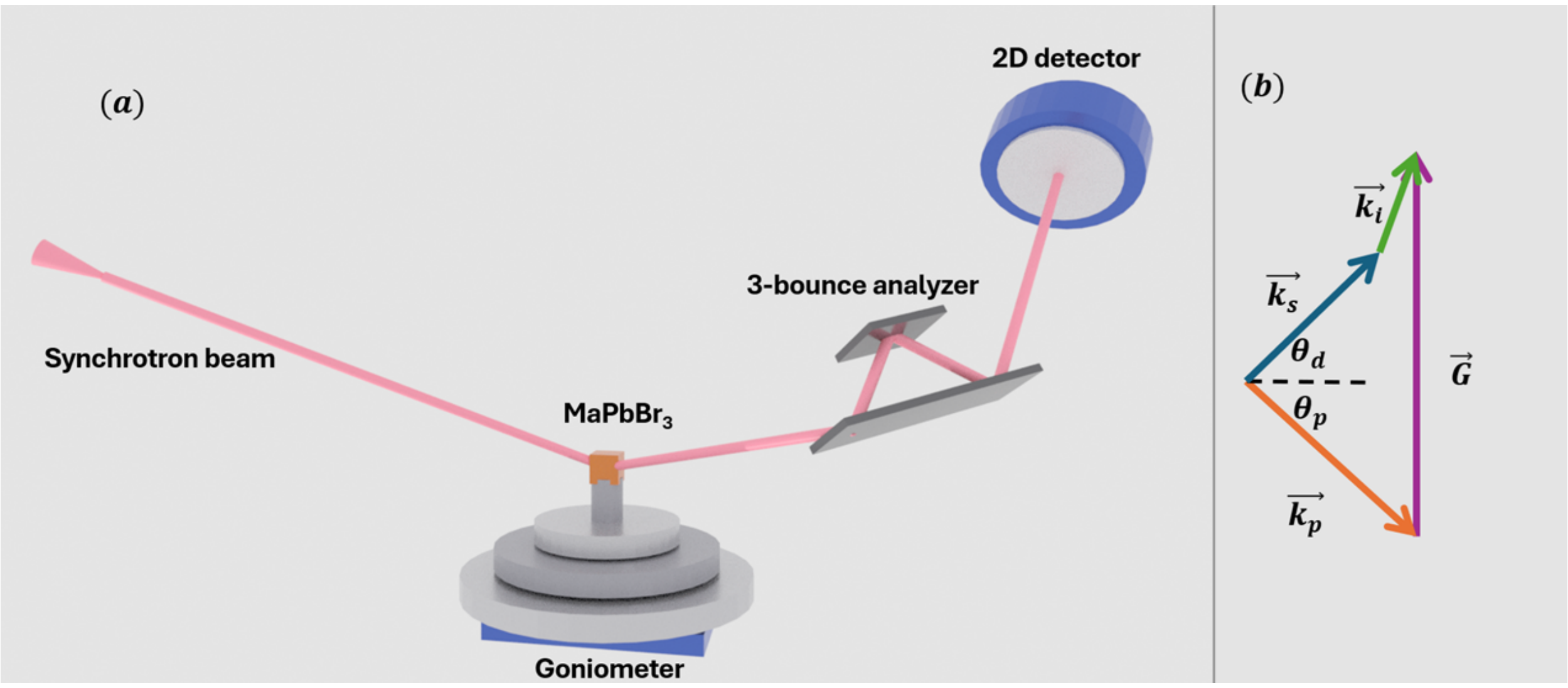

**Fig 1:** (a) Experimental geometry for X-ray PDC in $MAPbBr_3$. An incident 10 keV beam generates correlated signal and idler photons that satisfy energy and momentum conservation through a reciprocal lattice vector G of the (3,0,0) reflection. The weak PDC signal is spatially separated from the much stronger elastic scattering using a three-bounce Si(111) crystal analyzer - operated as a tunable narrow-band energy filter that suppresses elastic scattering - and is subsequently recorded on a two-dimensional detector. (b) Phase-matching scheme. The indices p, s, and i represent the pump, signal, and idler, respectively. $\vec{G}$ is the reciprocal lattice vector and the angles $\theta_p$ and $\theta_d$ are the angles between the atomic planes and the pump wave vector and between the signal photon to the detector, respectively.

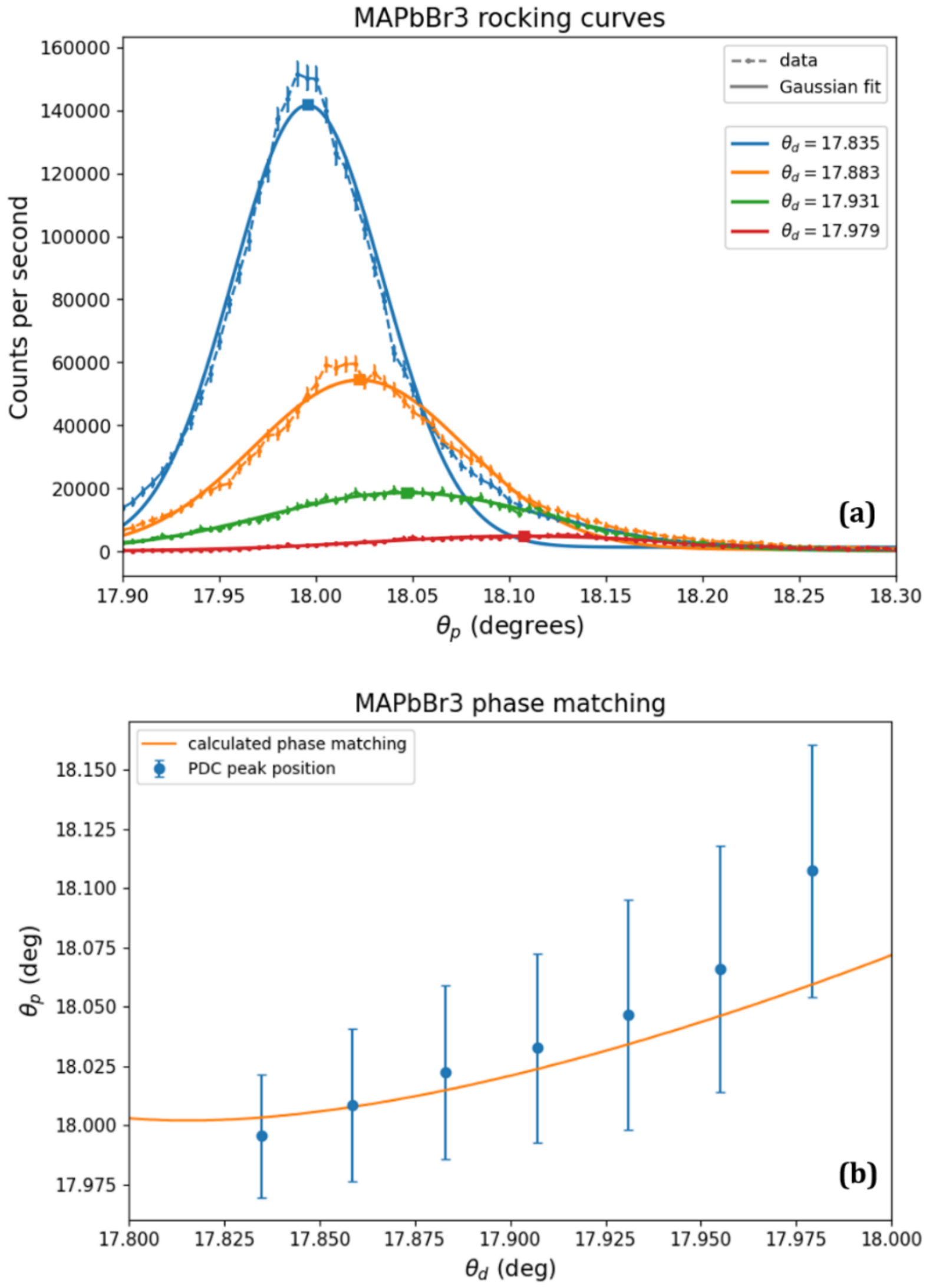

**Fig. 2**: (a) X-ray signal count rates dependence on the sample angle (rocking curves) for the (3 0 0) atomic planes at an idler energy of 40 eV, measured at different detector angles. Error bars assume Poisson statistics. The solid lines are Gaussian fits to the data. (b) Detector angle versus crystal angle at an idler energy of 40 eV, extracted from the rocking-curve peak positions for the (3 0 0) planes. The solid line shows the calculated phase-matching condition. The agreement between experiment and theory confirms the PDC origin of the signal. The bars indicate the angular width of each rocking curve at 80% of its maximum intensity, rather than the statistical uncertainty in the peak position.

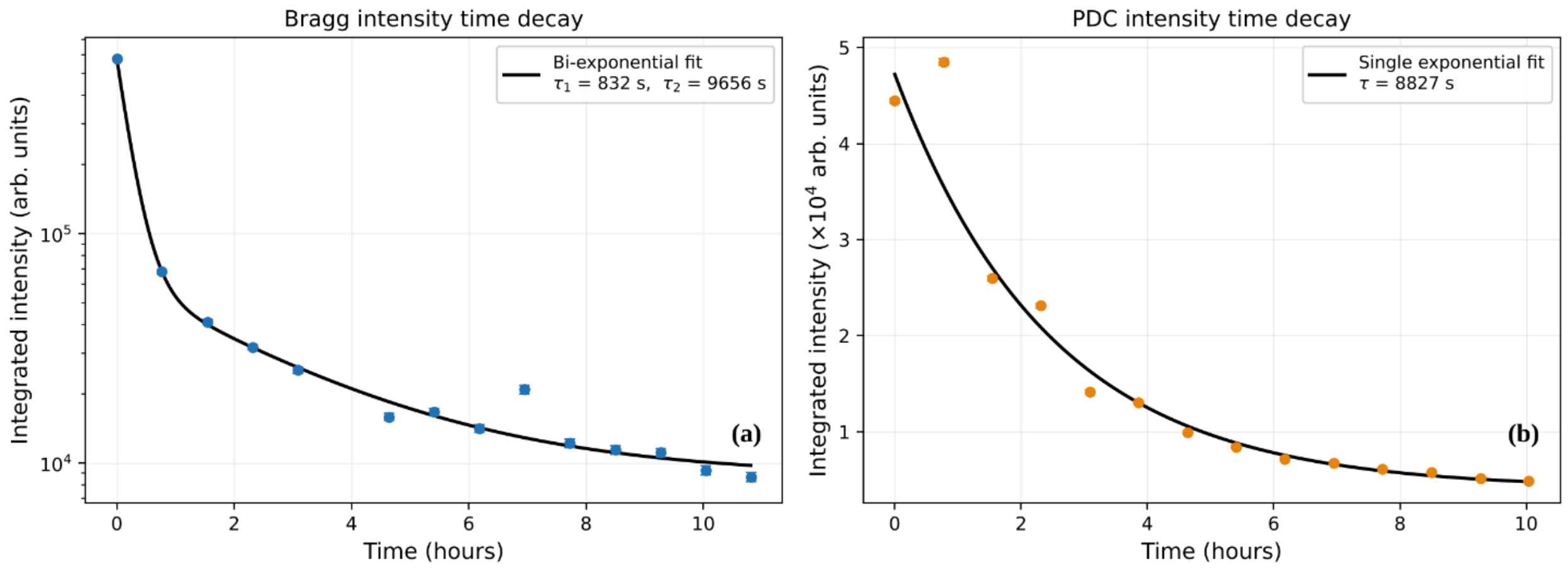

**Fig. 3**: Fitted time-resolved integrated intensity (3,0,0) (a) Bragg reflection (in log scale) and (b) PDC signal. The Bragg intensity is fitted to a bi-exponential decay to describe two distinct damage processes - a fast component ($\tau_1$ = 832 s) and a slow component ($\tau_2$ = 9656 s). The PDC intensity follows a single exponential ($\tau$ = 8827 s), in close agreement with the slow Bragg component. Error bars represent Poisson statistics. The Bragg point at about 7 hours deviates by about 5σ from the fit. Including it or excluding it shifts the fitted time constants by less than 1.5σ ($\tau_2$ = 9656 s vs. 8890 s), so we retain it in the fit.

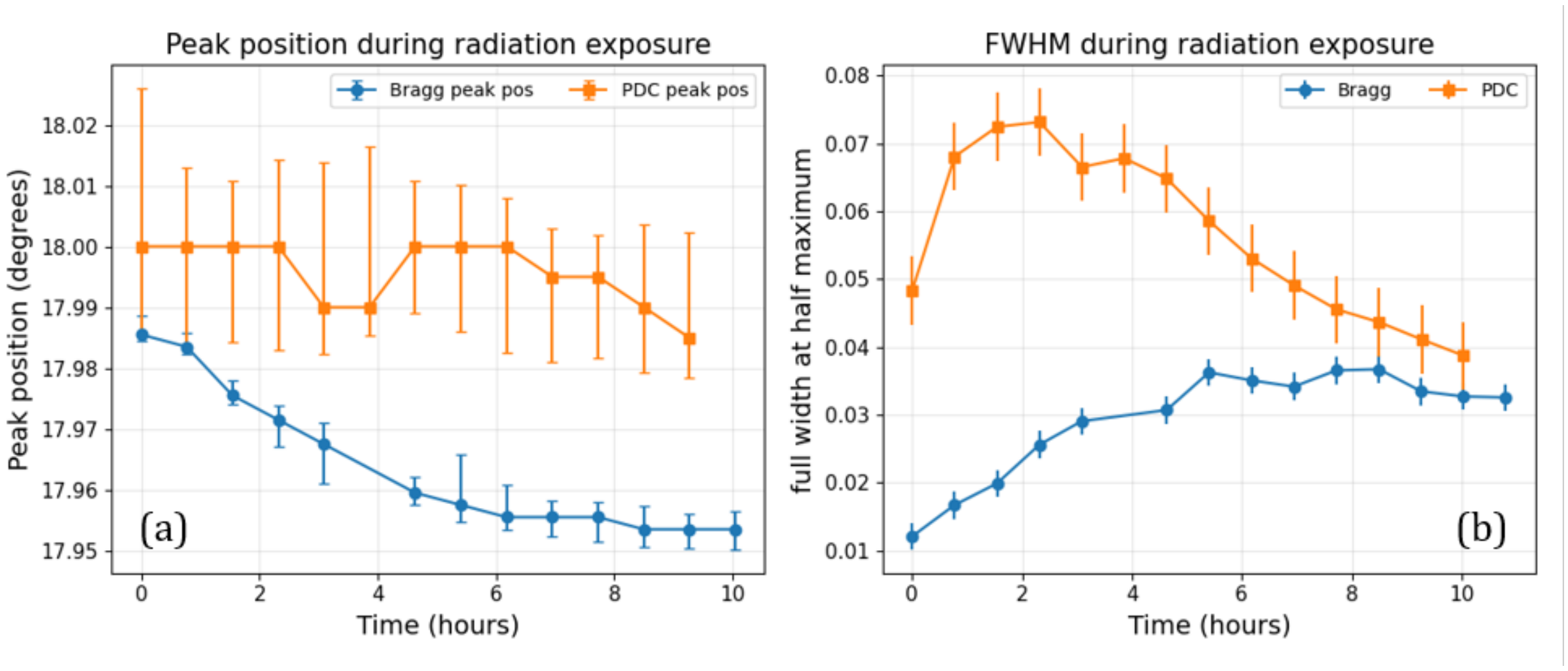

**Fig. 4**: (a) Peak position versus exposure time. The Bragg (3,0,0) peak shifts continuously to lower angle, from 17.985° to 17.953°. The PDC peak position remains essentially constant near 18.00° throughout the exposure. The bars indicate the angular width of each rocking curve at 80% of its maximum intensity, rather than the statistical uncertainty in the peak position. (b) FWHM versus exposure time. The Bragg FWHM increases monotonically from ~0.012° to a plateau near ~0.035°. The PDC FWHM rises to a maximum of ~0.073° within the first ~7000 s, then decreases to ~0.039°.

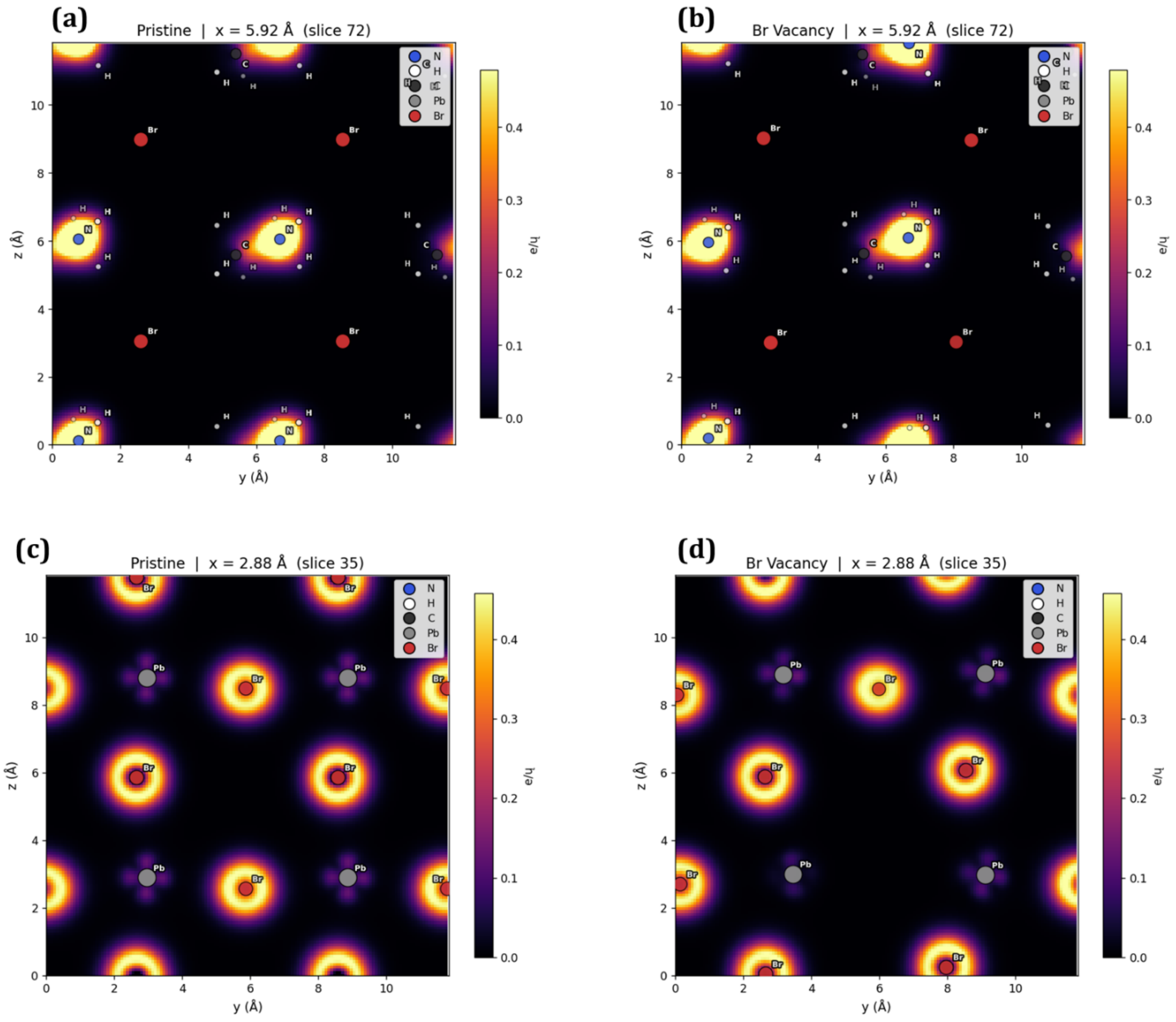

**Fig. 5:** Cross-sections of the integrated local density of states (ILDOS) computed by DFT for the 2×2×2 $MAPbBr_3$ supercell. (a,b) Partial charge density of the N 2s band in the yz-plane at x = 5.92 Å for the pristine (a) and bromide-vacancy (b) supercells. The density remains localized on the methylammonium cations and shows no visually discernible change upon vacancy formation. (c,d) Partial charge density of the Br 4s band in the yz-plane at x = 2.88 Å for the pristine (c) and vacancy (d) supercells. The missing bromide site in (d) is visible as an absent density lobe compared to (c). Atom positions within 1 Å of each slice plane are overlaid: N (blue), H (white), C (dark gray), Pb (gray), Br (red). Color scale indicates charge density in $e/a_0^3$.

| | $\tau(\mathbf{sec})$ | $D\left(\frac{cm^2}{sec}\right)$ |
|---|---|---|
| **Bragg** | $832 \pm 93$ | $(1.25 \pm 0.14) \cdot 10^{-8}$ |
| | $9656 \pm 2140$ | $(1.08 \pm 0.24) \cdot 10^{-9}$ |
| **PDC** | $8827 \pm 1070$ | $(1.18 \pm 0.14) \cdot 10^{-9}$ |

**Table 1:** Characteristic decay times and corresponding diffusion coefficients extracted from the time-dependent Bragg (3,0,0) and PDC intensity decays under X-ray exposure.

| | N 2s | | Br 4s | |
|---|---|---|---|---|
| | **Pristine** | **Vacancy** | **Pristine** | **Vacancy** |
| **Average density** ($\boldsymbol{e \cdot \text{Å}^{-3}}$) | 9.64e-3 | 9.64e-3 | 3.86e-2 | 3.74e-2 |
| $\lvert\boldsymbol{\rho_{3,0,0}}\rvert$ ($\boldsymbol{e \cdot \text{Å}^{-3}}$) | 2.82e-3 | 2.77e-3 | 4.52e-4 | 2.16e-4 |
| **Relative change:** $\frac{\lvert\rho_{3,0,0}\rvert_{pristine} - \lvert\rho_{3,0,0}\rvert_{vacacny}}{\lvert\rho_{3,0,0}\rvert_{pristine}}$ | 1.8% | | 52.2% | |

**Table 2:** Fourier component $|\rho_{3,0,0}|$ of the partial charge density for the N 2s and Br 4s energy windows, computed from the DFT ILDOS of the pristine and bromide-vacancy 2×2×2 $MAPbBr_3$ supercells. The G = (6,0,0) supercell component (corresponding to G = (3,0,0) of the primitive cell) was extracted by three-dimensional FFT of the real-space charge density. The average density and corresponding Fourier amplitudes are given in $e \cdot \text{Å}^{-3}$. The relative change quantifies the fractional modification of the diffraction-relevant Fourier component upon vacancy formation.